**Title:** Relating cognition to both brain structure and function: A systematic review of methods

**Authors:** Marta Czime Litwińczuk[a], Nelson Trujillo-Barreto[a], Nils Muhlert[a], Lauren Cloutman[a], Anna Woollams[a]

**Affiliations:** [a] Division of Neuroscience and Experimental Psychology, University of Manchester, UK

**Corresponding author:** Marta Czime Litwinczuk

**Email:** marta.litwinczuk@mancherster.ac.uk

**Postal Address:** Zochonis Building, Brunswick Street, Manchester M13 9GB, United Kingdom

**Funding**: This work was supported by the Biotechnology and Biological Sciences Research Council, UK (grant reference: BB/M011208/1).

**Competing interests statement:** No competing interests to declare.

**ABSTRACT:**

Cognitive neuroscience explores the mechanisms of cognition by studying its structural and functional brain correlates. Here, we report the first systematic review that assesses how information from structural and functional neuroimaging methods can be integrated to investigate the brain substrates of cognition. Web of Science and Scopus databases were searched for studies of healthy young adult populations that collected cognitive data, and structural and functional neuroimaging data. Five percent of screened studies met all inclusion criteria. Next, 54% of included studies related cognitive performance to brain structure and function without quantitative analysis of the relationship. Finally, 32% of studies formally integrated structural and functional brain data.

Overall, many studies consider either structural or functional neural correlates of cognition, and of those that consider both, they have rarely been integrated. We identified four emergent approaches to the characterisation of the relationship between brain structure, function and cognition; comparative, predictive, fusion and complementary. We discuss the insights provided each approach and how authors can select approaches to suit their research questions.

**MAIN TEXT:**

**1. INTRODUCTION**

Cognitive function and adaptive behaviour rely on structure and dynamics of large-scale neural networks (Friston, 2002). Early cognitive neuroscience separately assessed how properties and characteristics of brain structure and function might impact upon performance of cognitive tasks. Research employing the structural modality focused on studying physical properties of the brain, such as cytoarchitecture and neuronal integrity, whereas research employing functional approaches assessed characteristics of neuronal activity observed during performance of cognitive tasks and during rest (Rykhlevskaia et al., 2008). However, in recent years some attempts have been made to integrate the two approaches. Authors have begun to investigate how structure and function of the human brain relate to each other by assessing correspondence between findings from the two modalities (Johansen-Berg et al., 2004; Rykhlevskaia et al., 2008). This comparative approach produces a more complete understanding of healthy cognitive function across human lifespan (de Kwaasteniet et al., 2013; Guye et al., 2010; Hahn et al., 2013; Salami et al., 2016; van den Heuvel & Fornito, 2014; Minghui Wang et al., 2016).

There is a complex relationship between brain structure and function (Rykhlevskaia et al., 2008; Suárez et al., 2020). Independent labs have found a striking similarity between patterns of white matter fibres and functionally meaningful parcellations of the cortex (Greicius et al., 2009; Johansen-Berg et al., 2004; Jung et al., 2017). For example, the most central nodes of functional networks are directly and strongly connected by white matter tracts (Greicius et al., 2009). Some studies have focused on a temporal association of activity across remote regions, which is interpreted as interaction across these regions and commonly referred to as functional connectivity (Friston, 2002). Studies that compare patterns of structural white matter connectivity and functional connectivity have found moderate correspondence in structural and functional connectivity (Honey et al., 2009; Parker et al., 2003; Sporns et al., 2011; Wang et al., 2013). This indicates that there are many regions that are not directly connected, but still can show functional interactions (Ashourvan et al., 2019; Hagmann et al., 2008; Honey et al., 2009; Honey et al., 2010; Liao et al., 2015; Røge et al., 2017; Sun et al., 2017; Thomas et al., 2009). This implies that there are regions that are indirectly connected with each other and evidence demonstrates that accounting for indirect connections improves correspondence between structural and functional connectivity (Honey et al., 2009). This evidence illustrates that there is a complex and non-trivial relationship between brain structure and function. As a result of this complexity, it becomes challenging to interpret patterns of results in cognitive neuroimaging investigations when neural structure and function diverge, yet it is possible that divergence provides important information about the mechanisms involved.

Researchers have demonstrated that both regional and inter-regional relationships between brain structure and function can profoundly influence cognition in healthy and clinical populations. For example, one study investigated structural and functional differences across two aging groups with good and poor episodic memory (Persson et al., 2006). It was found that that severe decline in episodic memory was uniquely associated with reduced integrity of white matter in the anterior part of the corpus callosum and increased activity in right prefrontal cortex during episodic encoding. It was argued that the unique activity in right frontal regions observed for the older group with memory impairment may have been a compensatory mechanism for the structural disruption. In another example, both SC and FC have both been found to be decreased in temporal lobe epilepsy patients as compared to controls (Liao et al., 2011; Zhang et al., 2011). Furthermore, similarity between SC and FC has been found to be decreased in people with epilepsy (Chiang et al., 2015). In particular, this decoupling was then modulated by duration of epilepsy and structural changes to individual regions, which were unique to patients with left versus right temporal lobe epilepsy. Unique patterns of disruption of coupling between brain structure and function have been reported in other aspects of aging including emotion processing, executive function, language, motor function inhibition (Ford & Kensinger, 2014; Hu et al., 2013; Mander et al., 2017; Ritchie et al., 2018; Sun et al., 2017), and clinical disorders including schizophrenia, depression autism, stroke, dementia and many others (Anderson et al., 2011; Carter et al., 2009; Cocchi et al., 2014; Hojjati et al., 2018; Min Wang et al., 2016; Weinstein et al., 2011). In addition, several regression studies suggest that the relationship between structure and function contributes unique variance to explanation of cognitive performance (Dhamala et al., 2021; Mansour et al., 2021; Robinson et al., 2021).

  To provide an overview of trends and developments within this research field, the present systematic review assesses how researchers have attempted to combine structural and functional brain imaging data in healthy adult populations. The review considers the findings to date, relevant methodological considerations, and outstanding areas that need to be addressed. Through this we hope to gain a better understanding of the state of the field and highlight the potential of combining structural and functional neuroimaging data.

## 2. METHODOLOGY

The present work has been conducted in accordance with the guidelines for systematic reviews (Moher et al., 2009). First, four research questions were formulated: i) how many articles have included neuroimaging data and analysis of both structure and function from healthy adults, ii) what proportion of articles identified by the first research question have also obtained and analysed cognition, iii) what proportion of articles identified by the second research question have quantitatively characterised the relationship between neural structure and function, iv) what methods of statistical analysis have been used to make the quantitative comparison between neural structure and function.

To answer these research questions, Web of Science and Scopus databases were searched on the 21st of October in 2021. The following terms were used to search across topics, titles, abstracts and keywords: human brain, neuroimaging, structural, functional. The following terms were explicitly excluded from the search as they imply clinical research: pathology, disease, syndrome, disorder, reviews. The following search string has been used in Web of Science: "((TS=(human brain AND neuroimaging AND structural AND functional)) NOT TS=(pathology OR disease OR syndrome OR disorder)) NOT TS=(review)". The following search string has been used in Scopus: [( TITLE-ABS-KEY ( human AND brain AND neuroimaging AND functional AND structural ) AND NOT TITLE-ABS-KEY ( pathology OR disease OR syndrome OR disorder ) AND NOT TITLE-ABS-KEY ( review ) )]. Only formally published, peer-reviewed literature was included and 'grey literature' was excluded. In-text inclusion criteria were set to produce a report that is most representative of cognitive neuroimaging research conducted on healthy adult population. The full list of selection criteria, including article form, data analysis, study design and populations, can be found in Appendix 1A. Papers were selected for this review with the following process: papers were identified from databases, duplicate articles were removed, titles and abstracts were screened, and finally in-text elimination was conducted. As part of in-text elimination process, the articles without cognitive outcome were excluded. Article selection process was conducted by MCL.

The following information was recorded during data collection: cognitive task, neuroimaging acquisition protocols and paradigms, neuroimaging data pre-processing, outcome measures of neuroimaging data, scales of neuroimaging analysis, methods of integrating information about brain structure and function.

## 3. RESULTS

### 3.1. Literature Search and study characteristics

The process of identification, screening and selection of studies presented in Figure 1 has been obtained from PRISMA guidelines (Moher et al., 2009). First, 1923 articles were identified during database search and 251 records duplicates were removed. Then, 1513 records were removed during screening of titles and abstracts. Next, 159 articles were submitted to full-text assessment of eligibility, which resulted with further elimination of 58 articles. Overall, the selection process resulted with 102 articles that were included in our literature review and accounted for 5% of initial search results. Figure 2 illustrates the number of identified papers for each year.

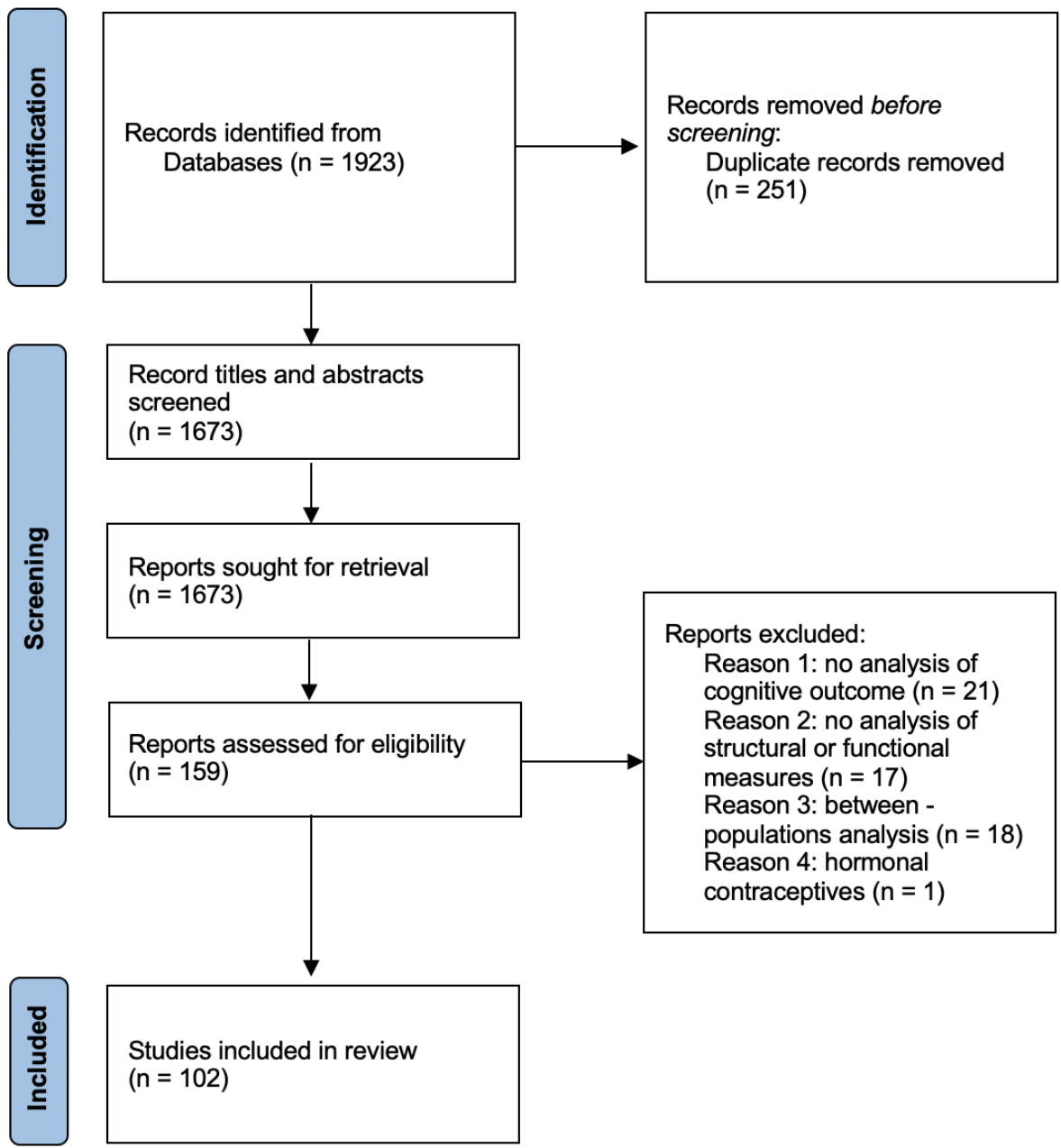

Figure 1 Flow diagram obtained from PRISMA guidelines for systematic reviews (Moher et al., 2009), it illustrates the literature search and literature selection process employed in present review.

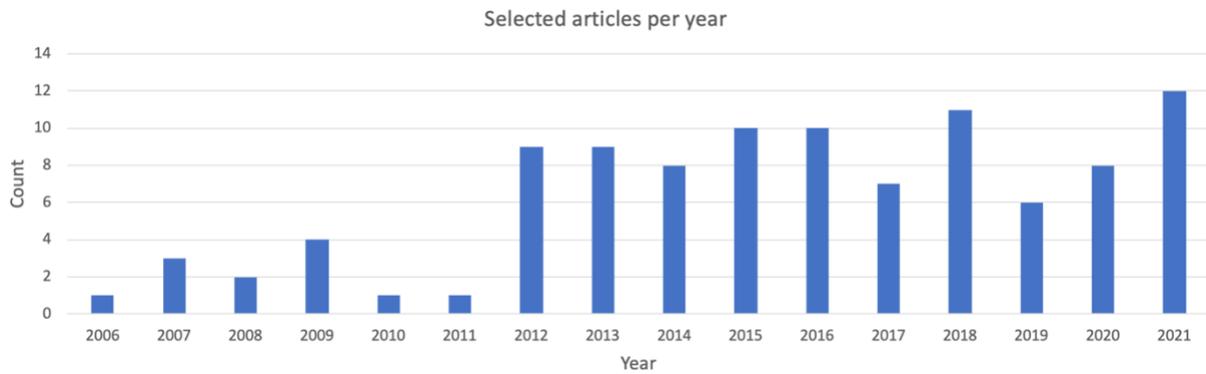

Figure 2 A time line represents count of selected publications for each year.

### 3.2. Cognitive Domains

The selected papers covered a wide variety of cognitive domains and their respective processes with many papers investigating multiple processes across many domains. Figure 3 illustrates the frequency of cognitive processes across the selected literature. Overall, the most investigated cognitive domain was language (featured in 21 articles), followed by memory (featured in 19 articles) and working memory (featured in 19 articles).

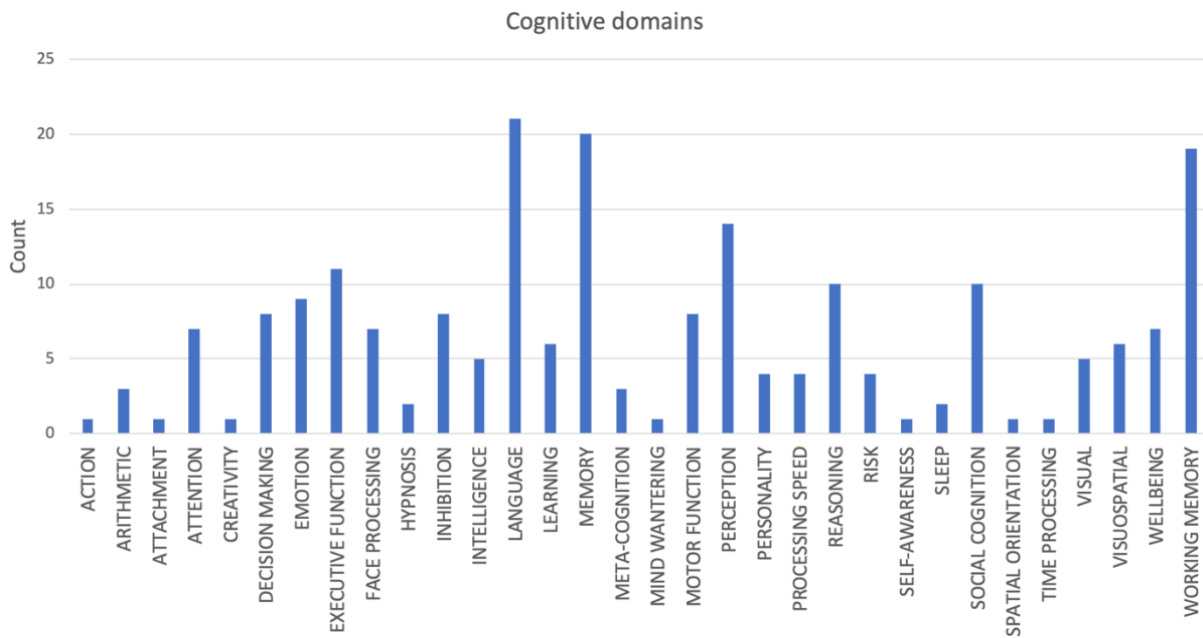

Figure 3 A distribution of cognitive processes across selected literature. Occurrence of cognitive processes was counted and if an article has investigated multiple cognitive process then the count was the fraction of all processes featured in the article.

### 3.3. Neuroimaging data and data analysis

The selected papers have shown an even balance between analysis of magnetic resonance imaging (MRI) and diffusion tensor imaging (DTI) (Figure 4), but functional MRI (fMRI) dominated the research field (Figure 5). The functional imaging was dominated by task paradigms, but resting state paradigm has also featured in many articles (Figure 6).

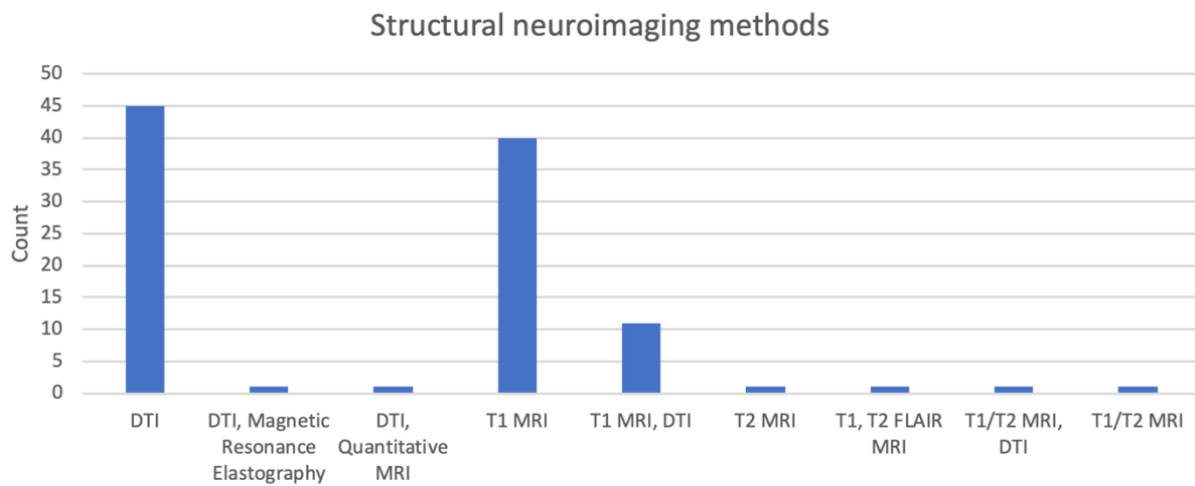

Figure 4 A distribution of structural neuroimaging methods across selected literature.

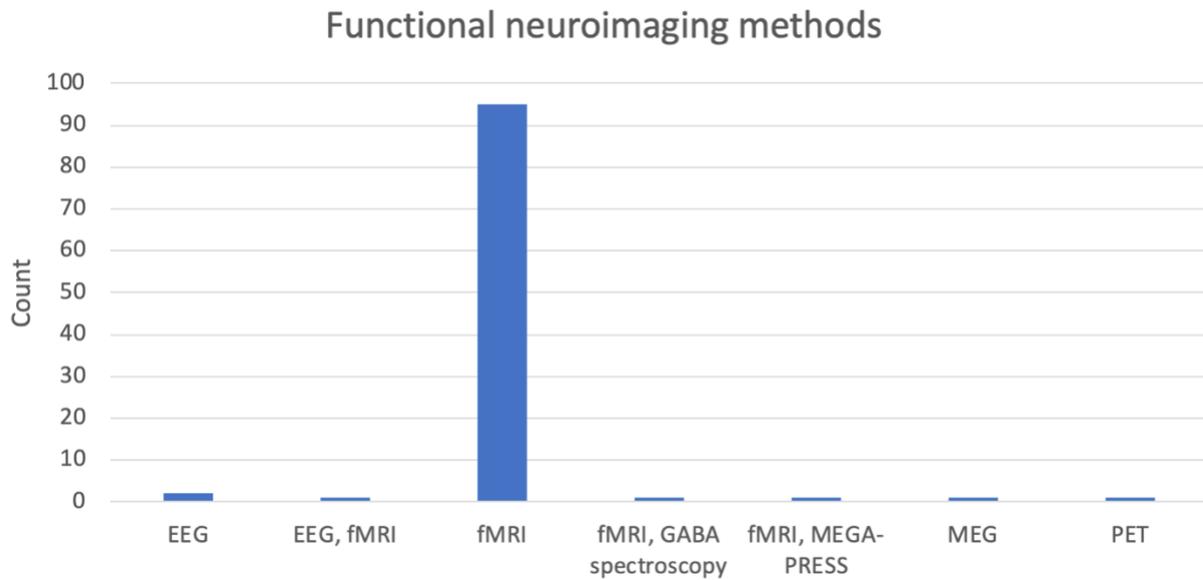

Figure 5 A distribution of functional neuroimaging methods across selected literature.

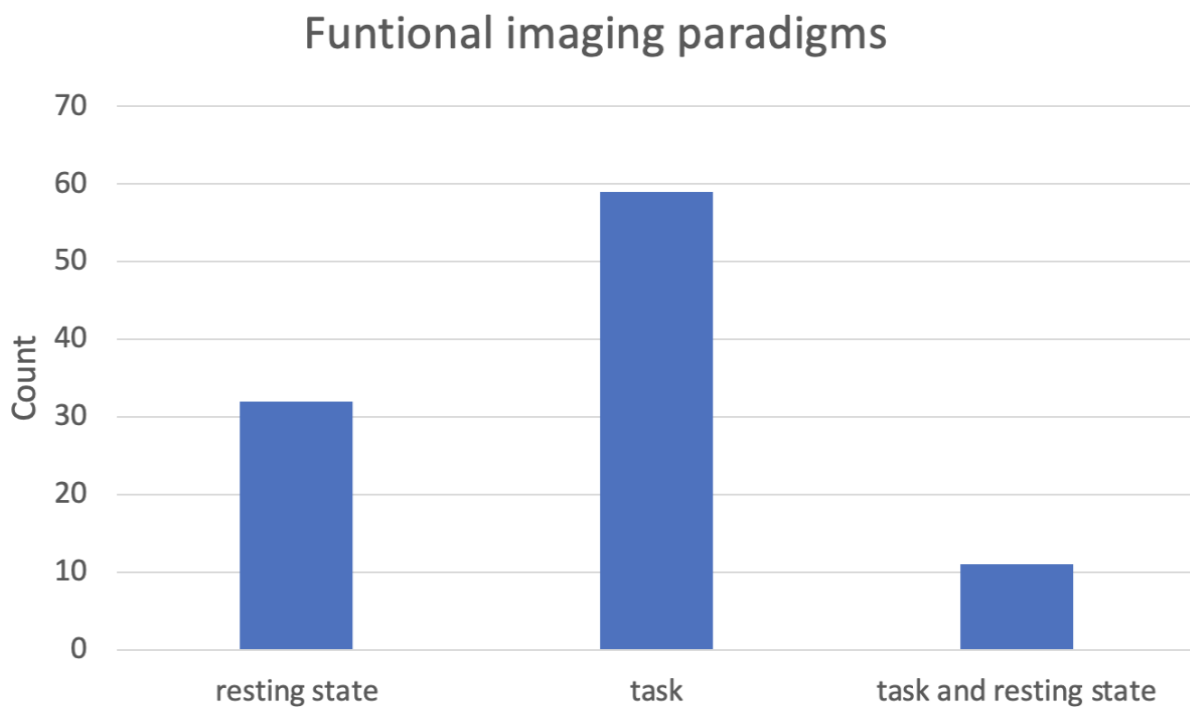

Figure 6 A distribution of functional neuroimaging paradigms across selected literature.

### 3.4. Methods of integrating structural and functional data

Studies were categorised by the types of inference they made about the relationship between brain structure and function: i) indirect, ii) semi-direct and iii) direct. First, studies that conducted separate analysis of brain structure and function without any quantitative evidence of their relationship were classified as making indirect inferences. Second, semi-direct inference referred to studies where authors have not provided statistical analyses of how brain structure and function relate to each other, but the analyses of each modality allowed some inference about how the modalities can relate. To illustrate, some studies have obtained common measures of properties of both structure and function but they have not tested how similar or different these measures are across modalities. In another example, some authors have used the topological location of results obtained for one modality to narrow down analysis of the other modality through informing location of ROI. Third, direct inference was defined as inference made with quantitative evidence to support the interpretation of results. The use of these methods of inference across studies is shown in Figure 7. Indirect inference was most common as it was used in 55 studies (53% of selected articles); flowed by direct inference in 29 studies (28%) and 4 of these latter studies used both direct and semi-direct inference (4%). 14 studies (14%) used semi-direct inference alone.

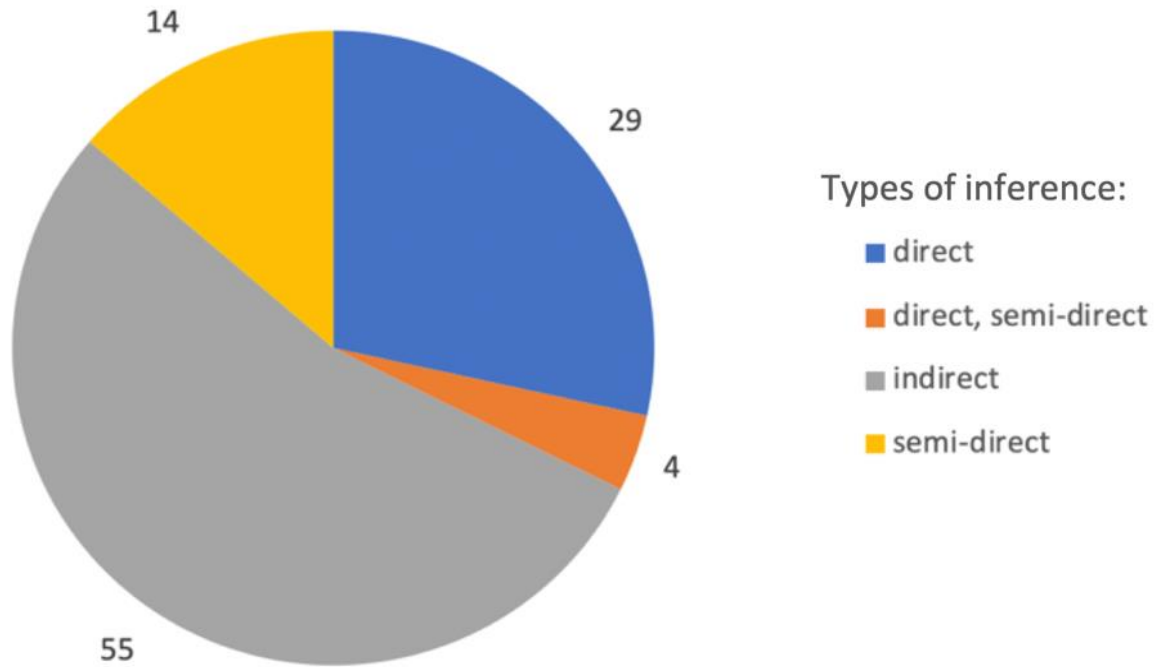

Figure 7 The percentage of structural and functional neuroimaging studies using different types of inference to understand their relationship. Direct inference was defined as inference backed by quantitative analysis of similarity in patterns observed between approaches. Semi-direct direct inference was defined as inference where analyses did not conduct direct inference, but some inference about how the modalities relate was still possible. Indirect inference was defined as inference made when neural structure and function were analysed separately.

The following methods of comparing brain structure and function were used by studies that only employed semi-direct inference. One study obtained volumetric measures in each neuroimaging dataset, one study assessed how much impact genetics had on each neuroimaging dataset, and one study obtained rich club coefficient in each neuroimaging dataset, and 11 studies used results from one modality to determine ROIs for the analysis of the other modality. As further illustrated in Figure 7, four studies combined direct and semi-direct approaches. Those studies used the results-driven ROI approach and conducted direct comparisons between structure and function through various methods, including correlation analyses and impact of structural information on models of effective connectivity.

The variety of direct inferences in quantitative analyses across studies are shown in Figure 8. Analysis of similarity (e.g. correlation, cosine) were the most common method of comparing brain structure and function. Next, many studies have produced various forms of joint models, where structural and functional data was used to predict cognitive outcomes. A handful of studies have either: used inferential statistics to assess the difference in characteristics of brain structure and function, attempted to predict functional models using structural information, assessed what ratio of functional connections had underlying direct structural connections, statistically assessed the overlap of results clusters for each modality, conducted data fusion such as independent component analysis, or assessed the effect of structural priors on functional model statistics.

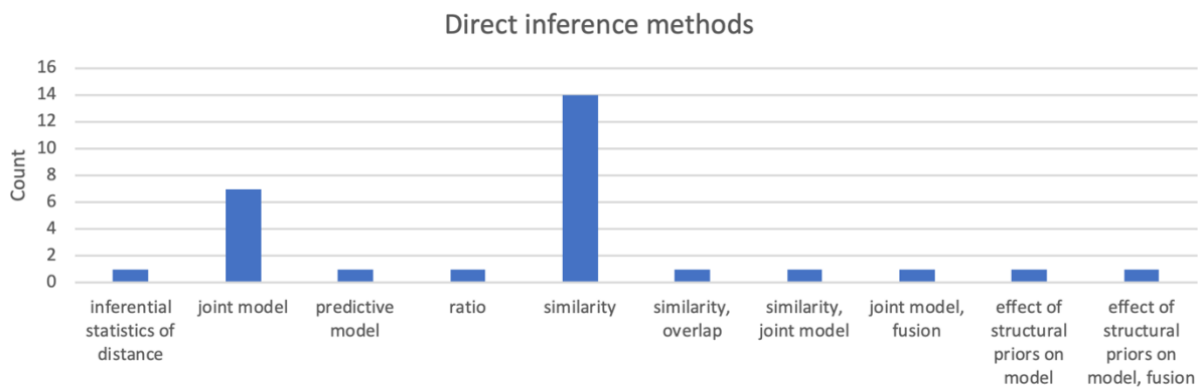

Figure 8 Methods used to relate structural and functional data in studies that made direct inferences.

**4. Discussion**

The relationship between brain structure and function may have profound consequences for understanding cognition (Ford & Kensinger, 2014; Hu et al., 2013; Mander et al., 2017; Ritchie et al., 2018; Sun et al., 2017). In this systematic review we determined how structural and functional neuroimaging methods have been integrated to study cognitive function and adaptive behaviour. A search was conducted across three databases in accordance with PRISMA guidelines (Moher et al., 2009). We assessed the prevalence of studies combining structural and functional neuroimaging data for explaining cognition and evaluated their choice of methods. The results demonstrate that there are to date relatively few studies attempting to combine structural and functional neuroimaging data and most studies use that do use indirect methods to infer the relationship between brain structure and function without formally relating these measures. Here, we consider what these findings mean for the field and how the shift towards direct inference with quantitative methods can lead to greater insight into how the structure and function of the brain combine to effect cognition.

First, this systematic review assessed the prevalence of studies quantitatively combining structural and functional data to explain cognition. Only 5% of the initial search results met the eligibility criteria, i.e. 102 out of 1923 studies examining links between structural and functional data and cognition in healthy adults. Investigations that address this subject have the potential to produce more complete understandings of healthy cognitive function, particularly given that combining brain structure and function information explains more variance in cognitive performance than either modality alone (Dhamala et al., 2021; Jiang et al., 2019; Robinson et al., 2021). These studies can also provide new insight, by determining e.g. causal interactions between regions (Sokolov, Zeidman, Erb, Ryvlin, Friston, et al., 2018; Sokolov, Zeidman, Erb, Ryvlin, Pavlova, et al.,

2018), or how new learning and training can result in neuroplasticity (Sun et al., 2016; Yang et al., 2019). Understanding relationships between brain structure, function and cognition can also provide insight into how these relationships breakdown in neurological and psychiatric disorders. The present review's selection criteria eliminated investigations of atypical populations, but some of the selected papers were then implemented in investigations into mechanisms of disease and recovery. To illustrate, Yang et al. (2019) investigated the effects of mindfulness training and were considered in later studies of general wellbeing of healthy populations (Tortella et al., 2021), improvements in cognitive function of diabetic patients (Alipor et al., 2019), and recovery from depression (van der Velden et al.). In another example, Jung et al. (2018) investigated how organisation of structural and functional connectivity relates to language, work that had direct implications in understanding language deficits in semantic variant primary progressive aphasia (Battistella et al., 2019) and temporal lobe epilepsy (Black, 2020). This illustrates the potential impact from a deeper understanding of how brain structure and function integrate.

      Having explored the prevalence of research on neural substrates of cognition in the literature, the present review considered methods for integrating across modalities. Three approaches to relating structure and function were observed: (i) a direct inference based on quantitative evidence, (ii) semi-direct inference based on closely related or similar processing steps, (iii) indirect inference based on separate analysis of the two approaches. Indirect inference was the most common approach. During this process, experimenters initially investigated how brain structure or function impacted cognitive function, next they inferred the degree of similarity between the two modalities. However, this process of investigation does not provide any information about how much additional explanatory

information is gained by analysing convergent structural and functional neural data. This means that research is focused on exploring the similarities and differences between brain structure and function, but it is still largely unknown how much the nature of this relationship impacts upon mechanisms of cognitive performance. Here, we recommend easily remedying this ambiguity by implementing measures of explained and residual variance in cognitive performance. By comparing the amount of explained variance in multimodal models, compared to single modality investigations, we can start to build better and perhaps more realistic models of cognition and its neural underpinnings. This will advance our understanding of which structural and functional properties are the most important and impactful on cognition, and how these may become vulnerable in the case of disease.

The present review also focused on methods used to make direct inferences. We can roughly place these methods under the umbrellas of four approaches; i) comparative approach, ii) predictive approach, iii) fusion approach and iv) complementary approach. First, the comparative approach assesses differences in characteristics of each network through measures of distance and significance testing. We have found that data was most commonly related across modalities with analysis of similarity such as correlation and quantification of results cluster overlap (Noonan et al., 2018). They were used to assess both linear and non-linear relationships between a variety of structural and functional properties. Structural properties included volumetrics values like cortical thickness and cortical surface area (Robinson et al., 2021), microstructure like grey and white matter directionality (e.g. fractional anisotropy) (Chavan et al., 2015), white matter connectivity like number of tract streamlines connecting region pairs (Sokolov, Zeidman, Erb, Ryvlin, Friston, et al., 2018), and symmetry scores assigned to reflect how symmetric these properties are across the two brain hemispheres (Josse et al., 2009). Functional properties have included properties of evoked activity like strength, count of activated voxels and laterality of activation (Zuo et al., 2016), and strength of FC as reflected by the correlation in signal intensities across remote regions (Robinson et al., 2021). Finally, inferential statistics have been used to assess if organisational characteristics of structural and functional networks significantly differ (Jung et al., 2018). The second approach of relating brain structure and function involves production of predictive models of cognition. Authors have employed multiple regression (Ford & Kensinger, 2014; Jung & Kim, 2020; Putnam et al., 2008), canonical correlation analysis (Han et al., 2020; Lerman-Sinkoff et al., 2017), partial least squares (Dzafic et al., 2019) and connectome-based predictive modelling (Jiang et al., 2019). One study started with sparsity-constrained principal component regression in each modality and the selected features of each modality were fed-forwards to a lasso analysis that integrated the models to a single model (Robinson et al., 2021). Another study produced predictions of cognitive performance separately using structural and functional connectivity and then calculated the average of the two predictions (Bajaj et al., 2021). Next, the fusion approach was observed where structural and functional information was fused prior to relating them to cognition. One such approach involved Independent Component Analysis, which was conducted on both connectivity sets and then canonical correlation analysis conducted on the resulting components (Lerman-Sinkoff et al., 2017). In another study, brain structure and function were fused, where structural connectivity produced a prior distribution of functional connectivity, which was then related to cognition (Xue et al., 2015). Finally, the complementary approach used information about structural characteristics to better understand the functional models of cognition. For example, one study evaluated how much additional variance is explained when structural priors are added to functional models (Kohno et al., 2017). Another group assessed if structural characteristics can be predictive of functional neural interactions observed between cognitive tasks (Chica et al., 2017). Finally, Sun et al. (2012) calculated the ratio of the effective connectivity observed during tasks with underlying direct structural white matter connections. It is important to note that some work may use several of these four general approaches. For example, partial least squares method can be used to fuse structural, functional and cognitive data, because it finds linear combination of predictors variables that covary with the response variable and projects all the information into a new space. Thus, we see that this method can both be used as a predictive and as fusion method depending on the kinds of research questions that the authors wish to address.

Each one of these approaches has their strengths in addressing specific research questions. The comparative approach allows us to understand what properties differ across structure and function. Consequently, the interpretation of results is easier and more meaningful as it explains what makes each modality unique and why unique patterns of results may be observed. However, it is important to remember that specific features of functional connectivity may be more difficult to be compared against features of brain structure. For example, there is evidence to demonstrate that there is indirect functional connectivity where functional connections can be observed in the absence of direct structural connections. In addition, functional connectivity can be negative, where a pair of remote regions shows a negative association in their signal strength. It is currently unclear how these unique functional characteristics should be compared against structural characteristics. Investigations that undertake complementary approach may be more suited to explore how much of brain function is related to brain structure, as they explore how prior information about brain structure can impact on the relationship between brain function and cognitive performance. The limitation of such investigations is however that they reflect little information about how neural function shapes neural structure, thus they provide a one-sided view of the relationship. Next, the predictive and fusion approaches have the capacity to effectively approach the multivariate nature of this research field. For example, Robinson et al. (2021) have started from a series of univariate regression models, composed of either structural or functional information, and implemented a stacked approach to eventually develop a combined model that is multivariate. In another example, Dhamala et al. (2021) have produced three regression models; using only structural connectivity, using only functional connectivity and using both. The challenge of such approaches remains that they may witness suppressor variables, where the variance in the response data accounted for by one variable may impact the beta weights of another variable (Lancaster, 1999). In addition, to our knowledge, so far research has focused on construction of linear models while interactions between structure and function have remained unexplored. Research into mediation effects in regression models will be necessary to more fully explore how the relationship between brain structure and function serves cognitive function. Thus we see that every approach can be used to answer slightly different questions about the relationship between brain structure and function, and authors can be creative in how they integrate several approaches to produce very refined models of cognition.

      This review also considered what data was acquired and how it was prepared for analysis. This highlighted a number of limitations. First, it became apparent that fMRI protocols have taken clear dominance over other functional imaging techniques in this research field. As mentioned in the introduction of this review, fMRI method suffers from low temporal resolution and is not a direct measure of neural activity. It is essential to

dedicate more research in the future to neuroimaging data with higher temporal resolution, such as EEG. This would allow more direct study of neural signals with millisecond precision. Consequently, signals that are not effectively reflected in the BOLD response could be studied, such as mismatch negativity which occurs 150 milliseconds following stimulus onset and is recognised as a marker of detection of stimulus irregularity (Näätänen, 1995). Second, there was a clear dominance of experimental protocols employing cognitive task performance over resting state fMRI. Resting state has been subjected to scrutiny and debate, as mental state and mental processes of the subjects are uncontrolled (Damoiseaux et al., 2006; Poldrack & Devlin, 2007; Smith et al., 2009). In contrast, task paradigms are carefully designed to engage and manipulate a cognitive process of interest and it is more clear what mental state was evoked in participants. However, resting state paradigms show moderate to high test-retest reliability and replicability across datasets and laboratories (Biswal et al., 2010; Buckner et al., 2009; Shehzad et al., 2009; Zuo, Di Martino, et al., 2010; Zuo, Kelly, et al., 2010). This means that resting state may allow easier comparison of results across independent research laboratories. Further, resting state produces consistent activation of a specific set of regions known as default mode network (Greicius et al., 2002). The function of this network has been related to cognitive functions, including but not limited to task switching, learning, and social cognition (McCormick & Telzer, 2018; Smith et al., 2018; Spunt et al., 2015). Further, its abnormal function has been implicated in many disorders such as dementia, schizophrenia, epilepsy, anxiety and depression and autism (Broyd et al., 2009). Third, many studies have employed correlation analysis as a method of relating brain structure and function. However, mediation analysis and partial correlation analyses were largely not employed. This is problematic, because it has been demonstrated that two regions can display functional connectivity in the absence of direct structural links between them and the similarity between functional and structural networks increases when indirect structural links are permitted in the analysis (Hagmann et al., 2008; Honey et al., 2009). This means that studies which ignore indirect links between regions may find less similarity between brain structure and function than studies that would account for those links.

To conclude, the present review was conducted to survey the prevalence of studies integrating brain structure and function for understanding cognition and detail the methods used in these analyses. Integrating structure and function and cognition is key for a full understanding of brain function and cognitive function through lifespan, disease and recovery. This review demonstrated that the relationship between brain structure and function and cognitive function is still largely underexplored. Inferences about the relationship between neural structure and function and cognitive function were indirect, semi-direct or direct, depending on what kind of evidence was used to support the interpretation of that relationship. Direct inference was not as common as indirect inference, and we have provided a brief discussion of available and previously used approaches to handling this multivariate analysis.

**REFERENCES:**

Alipor, A., Zare, H., Javanmard, G. H., & Mohammadi Garegozlo, R. (2019). Cognitive Remediation in Diabetics with Combining Mindfulness-based Relaxation and Trans-cranial Electrical Stimulation. *Iranian Journal of Health Psychology*, *2*(1), 31-46.

Anderson, J. S., Druzgal, T. J., Lopez-Larson, M., Jeong, E.-K., Desai, K., & Yurgelun-Todd, D. (2011). Network anticorrelations, global regression, and phase-shifted soft tissue correction. *Human Brain Mapping*, *32*(6), 919-934. https://doi.org/10.1002/hbm.21079

Ashourvan, A., Telesford, Q. K., Verstynen, T., Vettel, J. M., & Bassett, D. S. (2019). Multi-scale detection of hierarchical community architecture in structural and functional

brain networks. *PLoS One*, *14*(5), e0215520. https://doi.org/10.1371/journal.pone.0215520

Bajaj, S., Habeck, C., Razlighi, Q., & Stern, Y. (2021). Predictive utility of task-related functional connectivity vs. voxel activation. *PLoS One*, *16*(4). https://doi.org/10.1371/journal.pone.0249947

Battistella, G., Henry, M., Gesierich, B., Wilson, S. M., Borghesani, V., Shwe, W., Miller, Z., Deleon, J., Miller, B. L., & Jovicich, J. (2019). Differential intrinsic functional connectivity changes in semantic variant primary progressive aphasia. *NeuroImage: Clinical*, *22*, 101797.

Biswal, B. B., Mennes, M., Zuo, X. N., Gohel, S., Kelly, C., Smith, S. M., Beckmann, C. F., Adelstein, J. S., Buckner, R. L., Colcombe, S., Dogonowski, A. M., Ernst, M., Fair, D., Hampson, M., Hoptman, M. J., Hyde, J. S., Kiviniemi, V. J., Kotter, R., Li, S. J., . . . Milham, M. P. (2010). Toward discovery science of human brain function. *Proceedings of the National Academy of Sciences*, *107*(10), 4734-4739. https://doi.org/10.1073/pnas.0911855107

Black, J. (2020). *Cognitive Outcomes in Children with Temporal Lobe Epilepsy: Predictors of Academic Attainment* University of East London].

Broyd, S. J., Demanuele, C., Debener, S., Helps, S. K., James, C. J., & Sonuga-Barke, E. J. S. (2009). Default-mode brain dysfunction in mental disorders: A systematic review. *Neuroscience & Biobehavioral Reviews*, *33*(3), 279-296. https://doi.org/10.1016/j.neubiorev.2008.09.002

Buckner, R. L., Sepulcre, J., Talukdar, T., Krienen, F. M., Liu, H., Hedden, T., Andrews-Hanna, J. R., Sperling, R. A., & Johnson, K. A. (2009). Cortical Hubs Revealed by Intrinsic Functional Connectivity: Mapping, Assessment of Stability, and Relation to Alzheimer's Disease. *Journal of Neuroscience*, *29*(6), 1860-1873. https://doi.org/10.1523/jneurosci.5062-08.2009

Carter, A. R., Astafiev, S. V., Lang, C. E., Connor, L. T., Rengachary, J., Strube, M. J., Pope, D. L. W., Shulman, G. L., & Corbetta, M. (2009). Resting state inter-hemispheric fMRI connectivity predicts performance after stroke. *Annals of Neurology*, NA-NA. https://doi.org/10.1002/ana.21905

Chavan, C. F., Mouthon, M., Draganski, B., van der Zwaag, W., & Spierer, L. (2015). Differential patterns of functional and structural plasticity within and between inferior frontal gyri support training-induced improvements in inhibitory control proficiency. *Human Brain Mapping*, *36*(7), 2527-2543. https://doi.org/10.1002/hbm.22789

Chiang, S., Stern, J. M., Engel, J., & Haneef, Z. (2015). Structural–functional coupling changes in temporal lobe epilepsy. *Brain Research*, *1616*, 45-57. https://doi.org/10.1016/j.brainres.2015.04.052

Chica, A. B., Thiebaut de Schotten, M., Bartolomeo, P., & Paz-Alonso, P. M. (2017). White matter microstructure of attentional networks predicts attention and consciousness functional interactions. *Brain Structure and Function*, *223*(2), 653-668. https://doi.org/10.1007/s00429-017-1511-2

Cocchi, L., Harding, I. H., Lord, A., Pantelis, C., Yucel, M., & Zalesky, A. (2014). Disruption of structure–function coupling in the schizophrenia connectome. *NeuroImage: Clinical*, *4*, 779-787. https://doi.org/10.1016/j.nicl.2014.05.004

Damoiseaux, J. S., Rombouts, S. A. R. B., Barkhof, F., Scheltens, P., Stam, C. J., Smith, S. M., & Beckmann, C. F. (2006). Consistent resting-state networks across healthy subjects.

*Proceedings of the National Academy of Sciences*, *103*(37), 13848-13853. https://doi.org/10.1073/pnas.0601417103

de Kwaasteniet, B., Ruhe, E., Caan, M., Rive, M., Olabarriaga, S., Groefsema, M., Heesink, L., van Wingen, G., & Denys, D. (2013). Relation between structural and functional connectivity in major depressive disorder. *Biol Psychiatry*, *74*(1), 40-47. https://doi.org/10.1016/j.biopsych.2012.12.024

Dhamala, E., Jamison, K. W., Jaywant, A., Dennis, S., & Kuceyeski, A. (2021). Distinct functional and structural connections predict crystallised and fluid cognition in healthy adults. *Human Brain Mapping*, *42*(10), 3102-3118. https://doi.org/10.1002/hbm.25420

Dzafic, I., Oestreich, L., Martin, A. K., Mowry, B., & Burianová, H. (2019). Stria terminalis, amygdala, and temporoparietal junction networks facilitate efficient emotion processing under expectations. *Human Brain Mapping*, *40*(18), 5382-5396. https://doi.org/10.1002/hbm.24779

Ford, J. H., & Kensinger, E. A. (2014). The relation between structural and functional connectivity depends on age and on task goals. *Frontiers in Human Neuroscience*, *8*. https://doi.org/10.3389/fnhum.2014.00307

Friston, K. (2002). Functional integration and inference in the brain. *Progress in Neurobiology*, *68*(2), 113-143. https://doi.org/10.1016/s0301-0082(02)00076-x

Greicius, M. D., Krasnow, B., Reiss, A. L., & Menon, V. (2002). Functional connectivity in the resting brain: A network analysis of the default mode hypothesis. *Proceedings of the National Academy of Sciences*, *100*(1), 253-258. https://doi.org/10.1073/pnas.0135058100

Greicius, M. D., Supekar, K., Menon, V., & Dougherty, R. F. (2009). Resting-state functional connectivity reflects structural connectivity in the default mode network. *Cereb Cortex*, *19*(1), 72-78. https://doi.org/10.1093/cercor/bhn059

Guye, M., Bettus, G., Bartolomei, F., & Cozzone, P. J. (2010). Graph theoretical analysis of structural and functional connectivity MRI in normal and pathological brain networks. *MAGMA*, *23*(5-6), 409-421. https://doi.org/10.1007/s10334-010-0205-z

Hagmann, P., Cammoun, L., Gigandet, X., Meuli, R., Honey, C. J., Wedeen, V. J., & Sporns, O. (2008). Mapping the structural core of human cerebral cortex. *PLoS Biol*, *6*(7), e159. https://doi.org/10.1371/journal.pbio.0060159

Hahn, K., Myers, N., Prigarin, S., Rodenacker, K., Kurz, A., Förstl, H., Zimmer, C., Wohlschläger, A. M., & Sorg, C. (2013). Selectively and progressively disrupted structural connectivity of functional brain networks in Alzheimer's disease — Revealed by a novel framework to analyze edge distributions of networks detecting disruptions with strong statistical evidence. *NeuroImage*, *81*, 96-109. https://doi.org/10.1016/j.neuroimage.2013.05.011

Han, F., Gu, Y., Brown, G. L., Zhang, X., & Liu, X. (2020). Neuroimaging contrast across the cortical hierarchy is the feature maximally linked to behavior and demographics. *NeuroImage*, *215*. https://doi.org/10.1016/j.neuroimage.2020.116853

Hojjati, S. H., Ebrahimzadeh, A., Khazaee, A., & Babajani-Feremi, A. (2018). Predicting conversion from MCI to AD by integrating rs-fMRI and structural MRI. *Computers in Biology and Medicine*, *102*, 30-39. https://doi.org/10.1016/j.compbiomed.2018.09.004

Honey, C. J., Sporns, O., Cammoun, L., Gigandet, X., Thiran, J. P., Meuli, R., & Hagmann, P. (2009). Predicting human resting-state functional connectivity from structural connectivity. *Proceedings of the National Academy of Sciences*, *106*(6), 2035-2040. https://doi.org/10.1073/pnas.0811168106

Honey, C. J., Thivierge, J. P., & Sporns, O. (2010). Can structure predict function in the human brain? *NeuroImage*, *52*(3), 766-776. https://doi.org/10.1016/j.neuroimage.2010.01.071

Hu, S., Chao, H. H. A., Zhang, S., Ide, J. S., & Li, C.-S. R. (2013). Changes in cerebral morphometry and amplitude of low-frequency fluctuations of BOLD signals during healthy aging: correlation with inhibitory control. *Brain Structure and Function*, *219*(3), 983-994. https://doi.org/10.1007/s00429-013-0548-0

Jiang, R., Calhoun, V. D., Cui, Y., Qi, S., Zhuo, C., Li, J., Jung, R., Yang, J., Du, Y., Jiang, T., & Sui, J. (2019). Multimodal data revealed different neurobiological correlates of intelligence between males and females. *Brain Imaging and Behavior*, *14*(5), 1979-1993. https://doi.org/10.1007/s11682-019-00146-z

Johansen-Berg, H., Behrens, T. E. J., Robson, M. D., Drobnjak, I., Rushworth, M. F. S., Brady, J. M., Smith, S. M., Higham, D. J., & Matthews, P. M. (2004). Changes in connectivity profiles define functionally distinct regions in human medial frontal cortex. *Proceedings of the National Academy of Sciences*, *101*(36), 13335-13340. https://doi.org/10.1073/pnas.0403743101

Josse, G., Kherif, F., Flandin, G., Seghier, M. L., & Price, C. J. (2009). Predicting Language Lateralization from Gray Matter. *Journal of Neuroscience*, *29*(43), 13516-13523. https://doi.org/10.1523/jneurosci.1680-09.2009

Jung, J., Cloutman, L. L., Binney, R. J., & Lambon Ralph, M. A. (2017). The structural connectivity of higher order association cortices reflects human functional brain networks. *Cortex*, *97*, 221-239. https://doi.org/10.1016/j.cortex.2016.08.011

Jung, J., Visser, M., Binney, R. J., & Lambon Ralph, M. A. (2018). Establishing the cognitive signature of human brain networks derived from structural and functional connectivity. *Brain Structure and Function*, *223*(9), 4023-4038. https://doi.org/10.1007/s00429-018-1734-x

Jung, W. H., & Kim, H. (2020). Intrinsic Functional and Structural Brain Connectivity in Humans Predicts Individual Social Comparison Orientation. *Frontiers in Psychiatry*, *11*. https://doi.org/10.3389/fpsyt.2020.00809

Kohno, M., Morales, A. M., Guttman, Z., & London, E. D. (2017). A neural network that links brain function, white-matter structure and risky behavior. *NeuroImage*, *149*, 15-22. https://doi.org/10.1016/j.neuroimage.2017.01.058

Lancaster, B. P. (1999). Defining and Interpreting Suppressor Effects: Advantages and Limitations.

Lerman-Sinkoff, D. B., Sui, J., Rachakonda, S., Kandala, S., Calhoun, V. D., & Barch, D. M. (2017). Multimodal neural correlates of cognitive control in the Human Connectome Project. *Neuroimage*, *163*, 41-54. https://doi.org/10.1016/j.neuroimage.2017.08.081

Liao, W., Zhang, Z., Pan, Z., Mantini, D., Ding, J., Duan, X., Luo, C., Wang, Z., Tan, Q., Lu, G., & Chen, H. (2011). Default mode network abnormalities in mesial temporal lobe epilepsy: A study combining fMRI and DTI. *Human Brain Mapping*, *32*(6), 883-895. https://doi.org/10.1002/hbm.21076

Liao, X., Yuan, L., Zhao, T., Dai, Z., Shu, N., Xia, M., Yang, Y., Evans, A., & He, Y. (2015). Spontaneous functional network dynamics and associated structural substrates in the human brain. *Front Hum Neurosci*, *9*, 478. https://doi.org/10.3389/fnhum.2015.00478

Mander, B. A., Zhu, A. H., Lindquist, J. R., Villeneuve, S., Rao, V., Lu, B., Saletin, J. M., Ancoli-Israel, S., Jagust, W. J., & Walker, M. P. (2017). White Matter Structure in Older Adults Moderates the Benefit of Sleep Spindles on Motor Memory Consolidation. *The Journal of Neuroscience*, *37*(48), 11675-11687. https://doi.org/10.1523/jneurosci.3033-16.2017

Mansour, S. L., Tian, Y., Yeo, B. T. T., Cropley, V., & Zalesky, A. (2021). High-resolution connectomic fingerprints: Mapping neural identity and behavior. *NeuroImage*, *229*. https://doi.org/10.1016/j.neuroimage.2020.117695

McCormick, E. M., & Telzer, E. H. (2018). Contributions of default mode network stability and deactivation to adolescent task engagement. *Scientific Reports*, *8*(1). https://doi.org/10.1038/s41598-018-36269-4

Moher, D., Liberati, A., Tetzlaff, J., & Altman, D. G. (2009). Preferred Reporting Items for Systematic Reviews and Meta-Analyses: The PRISMA Statement. *PLoS Medicine*, *6*(7). https://doi.org/10.1371/journal.pmed.1000097

Näätänen, R. (1995). The Mismatch Negativity. *Ear and Hearing*, *16*(1), 6-18. https://doi.org/10.1097/00003446-199502000-00002

Noonan, M. P., Mars, R. B., Sallet, J., Dunbar, R. I. M., & Fellows, L. K. (2018). The structural and functional brain networks that support human social networks. *Behavioural Brain Research*, *355*, 12-23. https://doi.org/10.1016/j.bbr.2018.02.019

Parker, G. J. M., Haroon, H. A., & Wheeler-Kingshott, C. A. M. (2003). A framework for a streamline-based probabilistic index of connectivity (PICo) using a structural interpretation of MRI diffusion measurements. *Journal of Magnetic Resonance Imaging*, *18*(2), 242-254. https://doi.org/10.1002/jmri.10350

Persson, J., Nyberg, L., Lind, J., Larsson, A., Nilsson, L.-G., Ingvar, M., & Buckner, R. L. (2006). Structure–Function Correlates of Cognitive Decline in Aging. *Cerebral Cortex*, *16*(7), 907-915. https://doi.org/10.1093/cercor/bhj036

Poldrack, R. A., & Devlin, J. T. (2007). On the fundamental role of anatomy in functional imaging: Reply to commentaries on "In praise of tedious anatomy". *NeuroImage*, *37*(4), 1066-1068. https://doi.org/10.1016/j.neuroimage.2007.06.019

Putnam, M. C., Wig, G. S., Grafton, S. T., Kelley, W. M., & Gazzaniga, M. S. (2008). Structural Organization of the Corpus Callosum Predicts the Extent and Impact of Cortical Activity in the Nondominant Hemisphere. *Journal of Neuroscience*, *28*(11), 2912-2918. https://doi.org/10.1523/jneurosci.2295-07.2008

Ritchie, S. J., Cox, S. R., Shen, X., Lombardo, M. V., Reus, L. M., Alloza, C., Harris, M. A., Alderson, H. L., Hunter, S., Neilson, E., Liewald, D. C. M., Auyeung, B., Whalley, H. C., Lawrie, S. M., Gale, C. R., Bastin, M. E., McIntosh, A. M., & Deary, I. J. (2018). Sex Differences in the Adult Human Brain: Evidence from 5216 UK Biobank Participants. *Cerebral Cortex*, *28*(8), 2959-2975. https://doi.org/10.1093/cercor/bhy109

Robinson, E. C., Rasero, J., Sentis, A. I., Yeh, F.-C., & Verstynen, T. (2021). Integrating across neuroimaging modalities boosts prediction accuracy of cognitive ability. *PLoS Computational Biology*, *17*(3). https://doi.org/10.1371/journal.pcbi.1008347

Røge, R. E., Madsen, K. H., Schmidt, M. N., & Mørup, M. (2017). Infinite von Mises–Fisher Mixture Modeling of Whole Brain fMRI Data. *Neural Computation*, *29*(10), 2712-2741. https://doi.org/10.1162/neco_a_01000

Rykhlevskaia, E., Gratton, G., & Fabiani, M. (2008). Combining structural and functional neuroimaging data for studying brain connectivity: A review. *Psychophysiology*, *45*(2), 173-187. https://doi.org/10.1111/j.1469-8986.2007.00621.x

Salami, S. S., Obedian, E., Zimberg, S., & Olsson, C. A. (2016). Urinary quality of life outcomes in men who were treated with image-guided intensity-modulated radiation therapy for prostate cancer. *Adv Radiat Oncol*, *1*(4), 310-316. https://doi.org/10.1016/j.adro.2016.10.005

Shehzad, Z., Kelly, A. M. C., Reiss, P. T., Gee, D. G., Gotimer, K., Uddin, L. Q., Lee, S. H., Margulies, D. S., Roy, A. K., Biswal, B. B., Petkova, E., Castellanos, F. X., & Milham, M. P. (2009). The Resting Brain: Unconstrained yet Reliable. *Cerebral Cortex*, *19*(10), 2209-2229. https://doi.org/10.1093/cercor/bhn256

Smith, S. M., Fox, P. T., Miller, K. L., Glahn, D. C., Fox, P. M., Mackay, C. E., Filippini, N., Watkins, K. E., Toro, R., Laird, A. R., & Beckmann, C. F. (2009). Correspondence of the brain's functional architecture during activation and rest. *Proceedings of the National Academy of Sciences*, *106*(31), 13040-13045. https://doi.org/10.1073/pnas.0905267106

Smith, V., Mitchell, D. J., & Duncan, J. (2018). Role of the Default Mode Network in Cognitive Transitions. *Cerebral Cortex*, *28*(10), 3685-3696. https://doi.org/10.1093/cercor/bhy167

Sokolov, A. A., Zeidman, P., Erb, M., Ryvlin, P., Friston, K. J., & Pavlova, M. A. (2018). Structural and effective brain connectivity underlying biological motion detection. *Proceedings of the National Academy of Sciences*, *115*(51), E12034-E12042. https://doi.org/10.1073/pnas.1812859115

Sokolov, A. A., Zeidman, P., Erb, M., Ryvlin, P., Pavlova, M. A., & Friston, K. J. (2018). Linking structural and effective brain connectivity: structurally informed Parametric Empirical Bayes (si-PEB). *Brain Structure and Function*, *224*(1), 205-217. https://doi.org/10.1007/s00429-018-1760-8

Sporns, O., Pernice, V., Staude, B., Cardanobile, S., & Rotter, S. (2011). How Structure Determines Correlations in Neuronal Networks. *PLoS Computational Biology*, *7*(5). https://doi.org/10.1371/journal.pcbi.1002059

Spunt, R. P., Meyer, M. L., & Lieberman, M. D. (2015). The Default Mode of Human Brain Function Primes the Intentional Stance. *Journal of Cognitive Neuroscience*, *27*(6), 1116-1124. https://doi.org/10.1162/jocn_a_00785

Suárez, L. E., Markello, R. D., Betzel, R. F., & Misic, B. (2020). Linking Structure and Function in Macroscale Brain Networks. *Trends in Cognitive Sciences*, *24*(4), 302-315. https://doi.org/10.1016/j.tics.2020.01.008

Sun, J., Chen, Q., Zhang, Q., Li, Y., Li, H., Wei, D., Yang, W., & Qiu, J. (2016). Training your brain to be more creative: brain functional and structural changes induced by divergent thinking training. *Human Brain Mapping*, *37*(10), 3375-3387. https://doi.org/10.1002/hbm.23246

Sun, J., Hu, X., Huang, X., Liu, Y., Li, K., Li, X., Han, J., Guo, L., Liu, T., & Zhang, J. (2012). Inferring consistent functional interaction patterns from natural stimulus FMRI data. *NeuroImage*, *61*(4), 987-999. https://doi.org/10.1016/j.neuroimage.2012.01.142

Sun, Y., Li, J., Suckling, J., & Feng, L. (2017). Asymmetry of Hemispheric Network Topology Reveals Dissociable Processes between Functional and Structural Brain Connectome in Community-Living Elders. *Frontiers in Aging Neuroscience*, *9*. https://doi.org/10.3389/fnagi.2017.00361

Thomas, A. G., Marrett, S., Saad, Z. S., Ruff, D. A., Martin, A., & Bandettini, P. A. (2009). Functional but not structural changes associated with learning: An exploration of longitudinal Voxel-Based Morphometry (VBM). *NeuroImage*, *48*(1), 117-125. https://doi.org/10.1016/j.neuroimage.2009.05.097

Tortella, G. R., Seabra, A. B., Padrão, J., & Díaz-San Juan, R. (2021). Mindfulness and Other Simple Neuroscience-Based Proposals to Promote the Learning Performance and Mental Health of Students during the COVID-19 Pandemic. *Brain Sciences*, *11*(5). https://doi.org/10.3390/brainsci11050552

van den Heuvel, M. P., & Fornito, A. (2014). Brain Networks in Schizophrenia. *Neuropsychology Review*, *24*(1), 32-48. https://doi.org/10.1007/s11065-014-9248-7

van der Velden, A. M., Scholl, J., Elmholt45, E.-M., Fjorback, L., Harmer, C., Lazar, S., O'Toole, M., Smallwood, J., Roepstorff, A., & Kuyken, W. Mindfulness training changes brain dynamics during depressive rumination: A randomized controlled trial.

Wang, M., Jiang, S., Yuan, Y., Zhang, L., Ding, J., Wang, J., Zhang, J., Zhang, K., & Wang, J. (2016). Alterations of functional and structural connectivity of freezing of gait in Parkinson's disease. *Journal of Neurology*, *263*(8), 1583-1592. https://doi.org/10.1007/s00415-016-8174-4

Wang, M., Roussos, P., McKenzie, A., Zhou, X., Kajiwara, Y., Brennand, K. J., De Luca, G. C., Crary, J. F., Casaccia, P., Buxbaum, J. D., Ehrlich, M., Gandy, S., Goate, A., Katsel, P., Schadt, E., Haroutunian, V., & Zhang, B. (2016). Integrative network analysis of nineteen brain regions identifies molecular signatures and networks underlying selective regional vulnerability to Alzheimer's disease. *Genome Medicine*, *8*(1). https://doi.org/10.1186/s13073-016-0355-3

Wang, Z., Chen, Li M., Négyessy, L., Friedman, Robert M., Mishra, A., Gore, John C., & Roe, Anna W. (2013). The Relationship of Anatomical and Functional Connectivity to Resting-State Connectivity in Primate Somatosensory Cortex. *Neuron*, *78*(6), 1116-1126. https://doi.org/10.1016/j.neuron.2013.04.023

Weinstein, M., Ben-Sira, L., Levy, Y., Zachor, D. A., Itzhak, E. B., Artzi, M., Tarrasch, R., Eksteine, P. M., Hendler, T., & Bashat, D. B. (2011). Abnormal white matter integrity in young children with autism. *Human Brain Mapping*, *32*(4), 534-543. https://doi.org/10.1002/hbm.21042

Xue, W., Bowman, F. D., Pileggi, A. V., & Mayer, A. R. (2015). A multimodal approach for determining brain networks by jointly modeling functional and structural connectivity. *Frontiers in Computational Neuroscience*, *9*. https://doi.org/10.3389/fncom.2015.00022

Yang, C.-C., Barrós-Loscertales, A., Li, M., Pinazo, D., Borchardt, V., Ávila, C., & Walter, M. (2019). Alterations in Brain Structure and Amplitude of Low-frequency after 8 weeks of Mindfulness Meditation Training in Meditation-Naïve Subjects. *Scientific Reports*, *9*(1). https://doi.org/10.1038/s41598-019-47470-4

Zhang, Z., Liao, W., Chen, H., Mantini, D., Ding, J.-R., Xu, Q., Wang, Z., Yuan, C., Chen, G., Jiao, Q., & Lu, G. (2011). Altered functional–structural coupling of large-scale brain networks in idiopathic generalized epilepsy. *Brain*, *134*(10), 2912-2928. https://doi.org/10.1093/brain/awr223

Zuo, X.-N., Di Martino, A., Kelly, C., Shehzad, Z. E., Gee, D. G., Klein, D. F., Castellanos, F. X., Biswal, B. B., & Milham, M. P. (2010). The oscillating brain: Complex and reliable. *NeuroImage*, *49*(2), 1432-1445. https://doi.org/10.1016/j.neuroimage.2009.09.037

Zuo, X.-N., Kelly, C., Adelstein, J. S., Klein, D. F., Castellanos, F. X., & Milham, M. P. (2010). Reliable intrinsic connectivity networks: Test–retest evaluation using ICA and dual

-
regression approach. *NeuroImage*, *49*(3), 2163-2177. https://doi.org/10.1016/j.neuroimage.2009.10.080

Zuo, X.-N., Vassal, F., Schneider, F., Boutet, C., Jean, B., Sontheimer, A., & Lemaire, J.-J. (2016). Combined DTI Tractography and Functional MRI Study of the Language Connectome in Healthy Volunteers: Extensive Mapping of White Matter Fascicles and Cortical Activations. *PLoS One*, *11*(3). https://doi.org/10.1371/journal.pone.0152614